# Quantum and classical ratchet motions of vortices in a 2D trigonal superconductor

Yuki M. Itahashi[1]†, Yu Saito[1]†, Toshiya Ideue*[1], Tsutomu Nojima[2], Yoshihiro Iwasa[1,3]*

[1] *Quantum-Phase Electronics Center (QPEC) and Department of Applied Physics, The University of Tokyo, Tokyo 113-8656, Japan*

[2] *Institute for Materials Research, Tohoku University, Sendai 980-8577, Japan*

[3] *RIKEN Center for Emergent Matter Science (CEMS), Wako 351-0198, Japan*

*Corresponding author: ideue@ap.t.u-tokyo.ac.jp, iwasa@ap.t.u-tokyo.ac.jp

†These authors contributed equally to this work.

**Abstract**

**Dynamical behavior of vortices plays central roles in the quantum phenomena of two-dimensional (2D) superconductors. Quantum metallic state, for example, showing an anomalous temperature-independent resistive state down to low-temperatures, has been a common subject in recently developed 2D crystalline superconductors, whose microscopic origin is still under debate. Here, we unveil a new aspect of the vortex dynamics in a noncentrosymmetric 2D crystalline superconductor of $MoS_2$ through the nonreciprocal transport measurement. The second harmonic resistance $R^{2\omega}$ at low temperature with high current indicates the classical vortex flow accompanying the ratchet motion. Furthermore, we found that $R^{2\omega}$ is substantially suppressed in the quantum metallic state with low current region, allowing identification of the quantum and classical ratchet motions of vortices by the magnitude of the second harmonic generation. This suggests that nonreciprocal transport measurement can be a powerful tool to probe the vortex dynamics in noncentrosymmetric 2D superconductors.**

A variety of two-dimensional (2D) superconductors have emerged in the past decade, providing a novel materials' platform of unique physical phenomena[1–5]. Among them, quantum phases and their transitions in the vortex states are of particular interest[6–8]. In conventional amorphous metallic films, a superconductor-insulator transition occurring at a single critical point in the zero temperature limit is one of the well-known quantum phenomena[9,10]. Such quantum critical behavior is predominantly controlled by the disorder of films, which are inevitably enhanced by the reduction of film thickness, and hence well described by the scaling theory in the framework of the dirty boson picture[10].

On the other hand, a new route has been desired to approach the quantum phase transition with minimal disorder, for the comprehensive understanding of 2D superconductors. The recently emerging 2D superconductors based on single crystals[1,2] are one of such candidate materials, and in fact, some of them reveal a completely different physical picture from the dirty boson model[9]. Instead of superconducting and insulating states separated by the quantum critical point, gate-induced and exfoliated 2D superconductors display a broad metallic state dominating a magnetic field and temperature ($B$-$T$) phase diagram[6,7]. In this state, once the out-of-plane magnetic field is applied, the zero-resistance state is immediately destroyed into the temperature-independent finite resistive state even at low temperatures. Such a resistive state is observed not only in the 2D crystalline superconductors but also in amorphous thin film superconductors with weak pinning effect[8,11,12], though its origin is still under debate. In addition to the extrinsic heating by the environmental noise[13], several intrinsic mechanisms such as quantum collective creep of vortices[11,14] and Bose metal[7,15,16] are being discussed. However, it is technically difficult to distinguish such various vortex states by direct microscopic observation especially in nanometer-thick flake crystals, which hinders non-transport probes.

As a new probe of vortex states, we introduce, in this study, nonreciprocal phenomena, which is sensitively probed by the second harmonic magnetoresistance. It is generally known that, in a noncentrosymmetric system, the electrical resistance becomes dependent on the current direction, when the time reversal symmetry is broken[17]. Although nonreciprocal transport was originally studied in artificial helical structures or interfaces[17,18], it is nowadays applied to various noncentrosymmetric crystalline systems[19,20], including superconductors without inversion symmetry[21,22]. Among them, ion-gated $MoS_2$, considered to be a trigonal 2D superconductor (Fig. 1a), can be an ideal platform for investigating the vortex dynamics through nonreciprocal transport.

Here, we report a study on nonreciprocal charge transport, which sensitively probes the vortex ratchet motion reflecting lattice symmetry, in ion-gated $MoS_2$ single crystals. We found that the anisotropic behavior in the second harmonic resistance that satisfies the selection rules for threefold symmetry of single layer $MoS_2$, indicating that the gate-induced superconductivity in multilayer $MoS_2$ obeys the symmetry of monolayer. More importantly, the nonreciprocal magnetoresistance observed down to the lowest temperature at a relatively high current density can be understood in terms of not the paraconductivity scenario near $T_c$[22] but the ratchet motion of vortices in the plastic flow regime, which is caused by the asymmetric restraining force with threefold symmetry. Furthermore, we found that the second harmonic resistance signal at a low current is substantially suppressed below a certain magnetic field, being consistent with the quantum creep (tunneling) picture of vortices in noncentrosymmetric media. The present result indicates that nonreciprocal transport serves as an effective probe to differentiate the quantum and classical nature of vortex dynamics in noncentrosymmetric superconductors.

**Results**

**Sample fabrications and transport properties in MoS$_2$-EDLTs** We prepared three EDLT samples (Fig. 1b, see Methods) with different configurations using two different current directions: parallel to the zigzag direction (configuration A; sample 1 and 3) and parallel to the armchair direction (configuration B; sample 2), as shown in Fig. 2b. Carrier densities of sample 1, 2 and 3 are $1.2 \times 10^{14}$ cm$^{-2}$, $1.8 \times 10^{14}$ cm$^{-2}$ and $1.3 \times 10^{14}$ cm$^{-2}$, respectively, which were estimated by the Hall resistance measurements at 15 K. The crystal orientations of the MoS$_2$ flakes were determined from the shape of the edge according to a previous study[23].

We then performed the AC transport measurements (see Methods). First, we measured the first ($R^\omega$) harmonic signals of longitudinal resistance ($R_{xx}$) to show typical superconducting properties in the present system. Figure 1c shows $R_{xx}$ as a function of temperature $T$ for sample 1 (black) and sample 2 (orange) in zero magnetic field at a gate voltage of $V_G$ = 5 V. As $T$ decreases, $R_{xx}$ continuously decreases and suddenly goes to zero below 10 K. The $T_c$ values of samples 1 and 2, defined as the midpoints of the resistive transitions, are 8.8 and 6.8 K, respectively. Figure 1d shows $R_{xx}$-$T$ in sample 3 with $T_c$ = 8.3 K under various magnetic fields $B$ (0–9 T) at a source-drain current ($I$) of 0.5 µA. The drop of sheet resistance becomes broadened with increasing $B$, and the transition disappears near 9 T. Figure 1e shows an Arrhenius plot of $R_{xx}$ in sample 3 (the same data as those in Fig. 1d). At high temperatures near $T_c$, the data shows activated behavior indicating the thermal creep region[6,24,25], while at lower temperatures, the resistance deviates from the thermally activated behavior and shows the almost temperature-independent behavior. Figure 1f shows a vortex phase diagram based on the analyses in Fig. 1e for sample 3, where the upper critical field curve $B_{c2}(T)$ is derived from $B$ and $T$ with 90% of normal state resistance, which is confirmed to be close to the mean field value[24] and consistent with the Wertharmer–Helfand–Hohenberg (WHH) theory[26] at low temperature. Below $T_{\text{cross}}$, which is determined as temperatures at which $R_{xx}$ deviates from the activated behavior (white circles in Fig. 1e), the phase diagram is dominated by the quantum metallic state, in agreement with the previous study[24].

**Second harmonic measurements in 2D superconducting MoS$_2$** Next, we focus on the second ($R^{2\omega}$) harmonic signals of longitudinal ($R_{xx}$) and transverse ($R_{xy}$) resistance. Figures 2b and c (d and e) show the $R^{\omega}$ and $R^{2\omega}$ harmonic signals as a function of $B$, respectively, at $T$ = 2 K and $I$ = 13 µA (15 µA) in configuration A (B). As $B$ increases, the $R^{2\omega}$ signal ($R_{xx}$ in Fig. 2c and $R_{xy}$ in Fig. 2e) observed in both configurations first increases, then reaches a peak and finally decreases to almost zero. This peak behavior of $R^{2\omega}$ in the $B$ region with small absolute values of $R^{\omega}$ indicates that the nonreciprocal signals are enhanced in the superconducting state, as shown in the previous study[22]. In addition, we note that in the configuration A (B), only the $R^{2\omega}$ component of $R_{xx}$ ($R_{xy}$) is observed without discernible signal in the other component. The $B$-linear background of $R^{2\omega}$ component observed in sample 2 (Fig. 2e) is possibly attributed to the second harmonic component of $I$, which inevitably appears because of geometrical unbalances in the source–drain electrodes (see Supplementary Materials Section II). These directional dependent presence or absence of the second harmonic signals is consistent with the threefold symmetry of the crystal structure, reflecting the high crystallinity of the present 2D superconducting system. Also, it is stressed that the symmetry of the gate-induced superconductivity in multilayer MoS$_2$ is identical to that of the monolayer MoS$_2$. The observation of reduced symmetry by the gate electric field is consistent with the consideration in the previous paper[27,28].

Figures 3a and b show $R^{\omega}$ and $R^{2\omega}$ component of $R_{xx}$ ($R_{xx}^{\omega}$ and $R_{xx}^{2\omega}$), respectively, as a function of $B$ at various temperatures (2–10 K) in sample 1 with the configuration A for $I$ = 13 µA. We found that the amplitude of $R_{xx}^{2\omega}$ at the peak position ($R_{peak}^{2\omega}$) increases with decreasing temperature (Fig. 3c) and is indiscernible in $T > T_c$. This increase implies that not the paraconductivity but the vortex dynamics plays a crucial role in the enhancement of nonreciprocal transport because the transport far below $T_c$ and $B_{c2}$ is governed by vortex motion.

In order to elucidate the origin of nonreciprocal transport at lower temperatures far below $T_c$, we measured the $B$ dependence of $R_{xx}^{\omega}$ and $R_{xx}^{2\omega}$ (Figs. 3d and e) for various values of $I$ (5 - 45 μA) at 2 K in sample 1. As shown in Figs. 3d and e, in the low current region ($I = 5 – 18$ μA), there exists a region of $B$ where $R_{xx}^{\omega}$ is finite while $R_{xx}^{2\omega}$ is indiscernible below a certain magnetic field $B_1$ (white triangle). Above $B_1$, $R_{xx}^{2\omega}$ suddenly appears, showing a peak structure, and then decreases toward the second threshold magnetic field $B_2$ (black triangle). Above $B_2$, $R_{xx}^{2\omega}$ becomes very small with sign reversal. As $I$ increases, $B_1$ gradually decreases and becomes almost 0 T at higher current region ($I > 18$ μA). Fig. 3f displays $R_{peak}^{2\omega}$ as a function of $I$. The characteristic dome-like behavior is similar to that of vortex ratchet effect observed in thin films with artificial asymmetric potentials[29–31], where the rectified AC-driven vortices produce a DC electric field showing a dome-like dependence on the AC current. The decrease of the rectification effect at high $B$ is due to the effective weakening of the pinning potentials by increasing the AC current. In addition, according to a theoretical paper, nonreciprocal transport occurs even in the plastic vortex flow regime, where the vortices move among the pinning potentials as classical particles, in two-dimensional noncentrosymmetric superconductors[32]. Therefore, in the present case of $MoS_2$, we propose that asymmetric pinning potentials, which might be caused by defects of the 2D crystals such as sulfur vacancies, is the origin of an inequivalent motion (ratchet effect) of plastic vortex flow to the current directions. The pinning potentials are likely of trigonal symmetry, reflecting the intrinsic crystal structure. Although this pinning potential is believed to be weak in gate-induced superconductivity in $MoS_2$[24], it should be finite and affects the vortex motions sensitively.

For further understanding of the relation between vortex motion and nonreciprocal resistance at the low current region, we plot $R_{xx}^{\omega}$ and $R_{xx}^{2\omega}$ at $I = 5$ μA against $B$ in Fig. 4a and b, respectively, with schematic images of vortex motion in each magnetic field region in the inset. In this particular device, the zero-resistance (ZR) state is observed at finite magnetic field up to 0.25 T (orange triangle), above which $R_{xx}^{\omega}$ increases in a nonlinear manner with $B$. This region is assigned as the quantum metallic phase[6]. Above 1.25 T, the behavior of $R_{xx}^{\omega}$ changes from nonlinear to quasi-linear against $B$, being suggestive of a crossover from the quantum metallic regime to the vortex flow regions[6]. Importantly, this crossover field of $R_{xx}^{\omega}$ is close to the onset field $B_1$ of the nonreciprocal signal $R_{xx}^{2\omega}$. This observation indicates that the ratchet effect is small in the quantum metallic region, whereas it is dramatically enhanced in the vortex flow region. In the plastic vortex flow region above $B_1$, the vortex motion is still affected by the asymmetric pinning potentials due to the trigonal lattice symmetry, and therefore the vortex motion in this state is regarded as a classical ratchet. On the other hand, in the quantum metallic region, the vortex motion is dominated by the quantum mechanical tunneling of vortices (more strictly, quantum collective creep). The suppression of ratchet effect in quantum mechanical motion and the enhanced ratchet effect in classical motions are consistent with the general physical understanding of the ratchet effect. Indeed, many theoretical arguments have been made on quantum and classical ratchet effects in various model systems[32–35]. The present results suggest that the vortex state in noncentrosymmetric 2D superconductors might be a model system for further investigation of quantum and classical ratchet effects.

With further increasing magnetic fields, the nonreciprocal signal $R_{xx}^{2\omega}$ is again suppressed. Because $B_2$ is close to the magnetic field where the resistance become a half of the normal state value, the vortex picture might not hold any more at $B > B_2$ and should be replaced by the fluctuation picture. Note that the $R_{xx}^{2\omega}$ signal sensitively captures the end of the vortex flow state to the fluctuation regime, which is obscure in the magnetoresistance $R_{xx}^{\omega}$.

We finally provide a dynamical phase diagram based on the nonreciprocal signals in Fig. 4c. Both the zero resistance (ZR) state and quantum metallic (QM) states are quickly suppressed by the current, and the plastic vortex flow region becomes dominant. With further increasing *I*, the plastic flow region is also suppressed, because superconductivity is almost killed by the current. The strong red color indicates large $R_{xx}^{2\omega}$ signal from the classical ratchet effect in the plastic vortex flow region, and takes a crucial role in determining this dynamical vortex phase diagram.

**Discussion**

We note that the electron temperature in the high current regions might be increased more or less in Fig. 4c. However, the plastic vortex flow region seems not to shrink abruptly with increasing current in our measurement range, implying the stable vortex dynamic state without quenching to normal state. Therefore, the heating effect on the nonreciprocal signal is not so significant if it exists and the dynamical phase diagram is not qualitatively altered. The suppression of nonreciprocal signals in the low current and low magnetic field region supports the scenario of quantum tunneling of vortices for the quantum metallic state[6], contrary to the recent study[13], which attributes a quantum metallic state to the heating effect by the extrinsic noise. Our results thus suggest that the nonreciprocal transport is a powerful tool to investigate the vortex dynamics in noncentrosymmetric 2D superconductors.

**Methods**

**Device fabrication.** Bulk 2*H*-MoS$_2$ single crystals were exfoliated into thin flakes by the Scotch-tape method and the flakes were transferred onto a Si/SiO$_2$ substrate. Thickness of exfoliated flakes were estimated to be 20 nm judging from optical contrasts. Also, steps which influence vortex motion were not found on the surface of flakes. Hall bar configuration was fabricated onto the flakes with Au (90 nm)/Cr (5 nm) electrodes. The pattern was fabricated with an electron beam lithography and electrodes were deposited with an evaporator. The direction of flakes was judged from their optical images. N,N-diethyl-N-(2-methoxyethyl)-N-methylammonium bis (trifluoromethylsulphonyl) imide (DEME-TFSI) was used as a gate medium.

**Transport measurements.** The temperature dependent resistance under a magnetic field was measured with a standard four-probe geometry in a Quantum Design Physical Property Measurement System (PPMS) combined with two kinds of AC lock-in amplifiers (Stanford Research Systems Model SR830 DSP and Signal Recovery Model 5210) with a frequency of 13 Hz. The gate voltage was supplied by a Keithley 2400 source meter at 220 K, which is just above the glass transition temperature of DEME-TFSI, under high vacuum (less than $10^{-4}$ Torr).

**Data availability.** The data that support the findings of this study are available from the corresponding author upon reasonable request.

**Acknowledgements**

We thank N. Nagaosa, S. Hoshino, R. Wakatsuki, K. Hamamoto, H. Ueki, M. Ohuchi, J. Laurienzo and Y. Nagai for fruitful discussions. Y.M.I. was supported by the Advanced Leading Graduate Course for Photon Science (ALPS). Y.S. was supported by the Japan Society for the Promotion of Science (JSPS) through a research fellowship for young scientists (Grant-in-Aid for JSPS Research Fellow, JSPS KAKENHI Grant Number JP15J07681). T.I. was supported by a Grant-in-Aid for Challenging Research (Exploratory) (JSPS KAKENHI Grant Number JP17K18748) and a Grant-in-Aid for Scientific Research on Innovative Areas "Topological Materials Science" (JSPS KAKENHI Grant Number 18H04216) from JSPS. This work was supported by a Grant-in-Aid for Specially Promoted Research (JSPS KAKENHI Grant Number JP25000003) from JSPS.

**Author contributions**

Y.M.I. and Y.S. contributed equally to this work. Y.S., T.I, and Y.I. conceived the research project. Y.S. designed the experiments. Y.M.I. fabricated samples. Y.M.I. and Y.S. performed the experiments and analyzed the data. All authors led the physical discussions and wrote the manuscript.

**Additional information**

Reprints and permissions information is available online at www.nature.com/reprints. Correspondence and requests for materials should be addressed to Y.S. and Y.I.

**Competing financial interests**

The authors declare no competing financial interests.

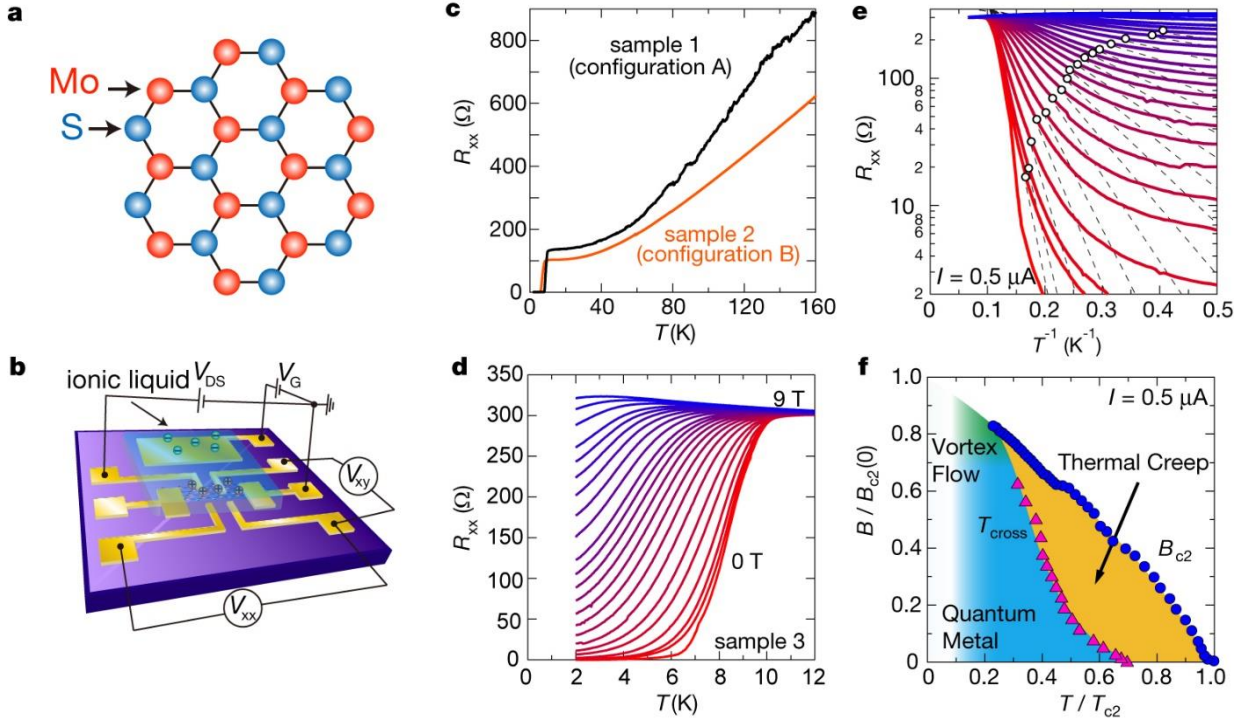

**Figure 1 | Gate-induced superconductivity and vortex phase diagram in MoS$_2$.** (**a**), Top view of trigonal crystal structure of MoS$_2$. (**b**), Schematic image of EDLT. (**c**), $R_{xx}$ as a function of $T$ in samples 1 (black line) and 2 (orange line) for the magnetic field at 0 T. Samples 1 and 2 correspond to configurations A and B, respectively. (**d**), $R_{xx}$ as a function of $T$ in sample 3 under various perpendicular magnetic fields. The magnetic field was varied in 0.05 T steps from 0 to 0.1 T, and in 0.2 T steps from 0.2 to 0.6 T, and in 0.3 T steps from 0.9 to 3 T, and in 0.5 T steps from 3.5 to 6 T, and in 1 T steps from 7 to 9 T. (**e**), Arrhenius plot of $R_{xx}$ in sample 3 for the same data as ones in Fig. 1d. The black dashed lines show the activated behavior of vortices described as $R_{xx} \propto \exp\left(-\dfrac{U}{k_B T}\right)$. White circles show boundary between thermal creep and quantum creep regime ($T_{\text{cross}}$). $T_{\text{cross}}$ is determined as temperatures at which $R_{xx}$ deviates from the activated behavior. (**f**), Vortex phase diagram in MoS$_2$-EDLT at a low current of 0.5 μA. The upper critical field curve $B_{c2}(T)$ or $B(T_{c2})$ (blue circles) is defined as $B$ or $T$ with 90 % of normal state resistance and $T_{\text{cross}}$ (pink triangles) determined from the Arrhenius plot as shown by the white circles in Fig. 1e. $B_{c2}(0) \sim 8$ T is defined by Werthamer-Helfand-Hohenberg formula [26], $B_{c2}(0) = 0.7 T_c \left(\dfrac{dB_{c2}}{dT}\right)\bigg|_{T_c}$. Orange and light blue region show thermal creep and quantum metal region, respectively.

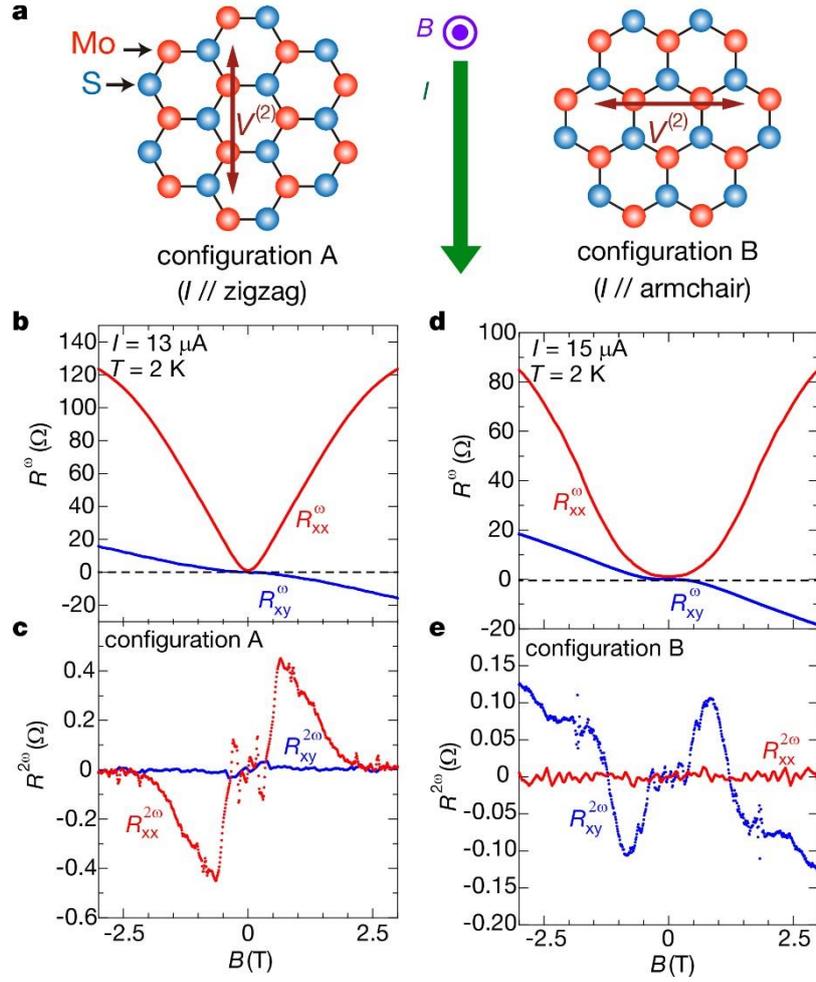

**Figure 2 | Selection rules of nonreciprocal transport reflecting threefold symmetry.** (**a**), Two kinds of device configurations depending on the crystal orientation of MoS$_2$. In configuration A (B), the applied current is parallel to the zigzag (armchair) direction, and a second harmonic signal of the longitudinal (transverse) component is expected. (**b**) and (**c**), Longitudinal ($R_{xx}$: red) and transverse ($R_{xy}$: blue) (b) first ($R^{\omega}$) and (c) second ($R^{2\omega}$) harmonic magnetoresistance when $T = 2$ K and $I = 13$ μA in configuration A, which is shown in the inset. $R_{xx}^{\omega}$ is symmetrized and others are antisymmetrized (see Supplemental Materials II in detail). (**d**) and (**e**), Longitudinal ($R_{xx}$: red) and transverse ($R_{xy}$: blue) (d) first ($R^{\omega}$) and (e) second ($R^{2\omega}$) harmonic magnetoresistance when $T = 2$ K and $I = 15$ μA in configuration B, which is shown in the inset. $R_{xx}^{\omega}$ is symmetrized and others are antisymmetrized.

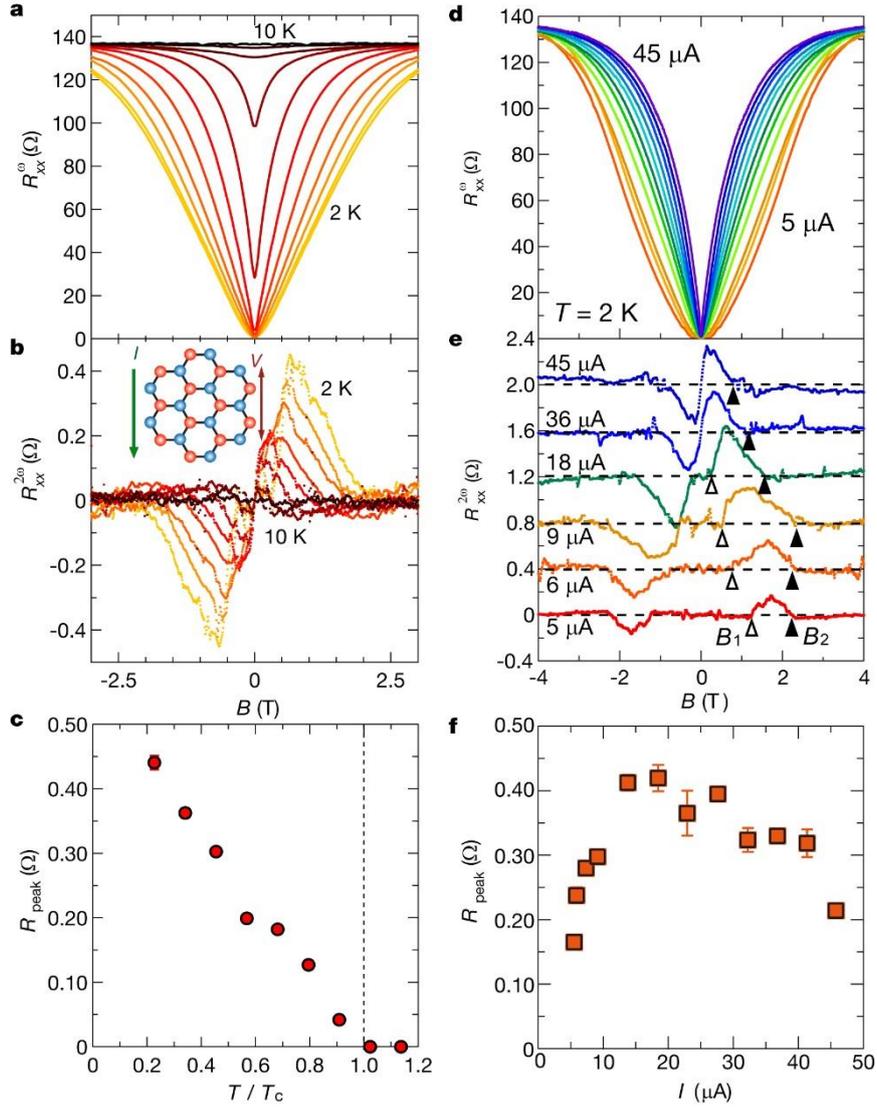

**Figure 3 | First and second harmonic magnetoresistance at various temperatures and currents for the longitudinal direction.** (**a**) and (**b**), Longitudinal first ($R_{xx}^{\omega}$) (a) and second ($R_{xx}^{2\omega}$) (b) harmonic magnetoresistance for temperatures varying in 0.5 K steps from 2 to 7 K, and 8, 9 and 10 K for $I = 13$ µA in configuration A, which is shown in the inset. $R_{xx}^{\omega}$ is symmetrized and $R_{xx}^{2\omega}$ is antisymmetrized. (**c**), Amplitude of second harmonic resistance at the peak position ($R_{\text{peak}}^{2\omega}$) extracted from Fig. 3b, as a function of temperature normalized with $T_c$ ($T/T_c$). (**d**), Longitudinal magnetoresistance ($R_{xx}^{\omega}$) for currents at 5, 6, 7, 9, 13, 18, 23, 28, 32, 36, 41 and 45 µA at $T = 2$ K in configuration A. (**e**), Second harmonic magnetoresistance ($R_{xx}^{2\omega}$) $R_{xx}^{\omega}$ for currents at 5, 6, 9, 18, 36 and 45 µA. Each curve is shifted vertically by 0.4 Ω for clarity. $R_{xx}^{\omega}$ is symmetrized and $R_{xx}^{2\omega}$ is antisymmetrized. $R_{xx}^{2\omega}$ is substantially enhanced between $B_1$ and $B_2$, as indicated by white and black triangles, respectively. (**f**), Amplitude of

$R_{xx}^{2\omega}$ at the peak position ($R_{peak}^{2\omega}$) as a function of current extracted from $R_{xx}^{2\omega}$ data. Errorbars indicate uncertainty of $R_{peak}^{2\omega}$.

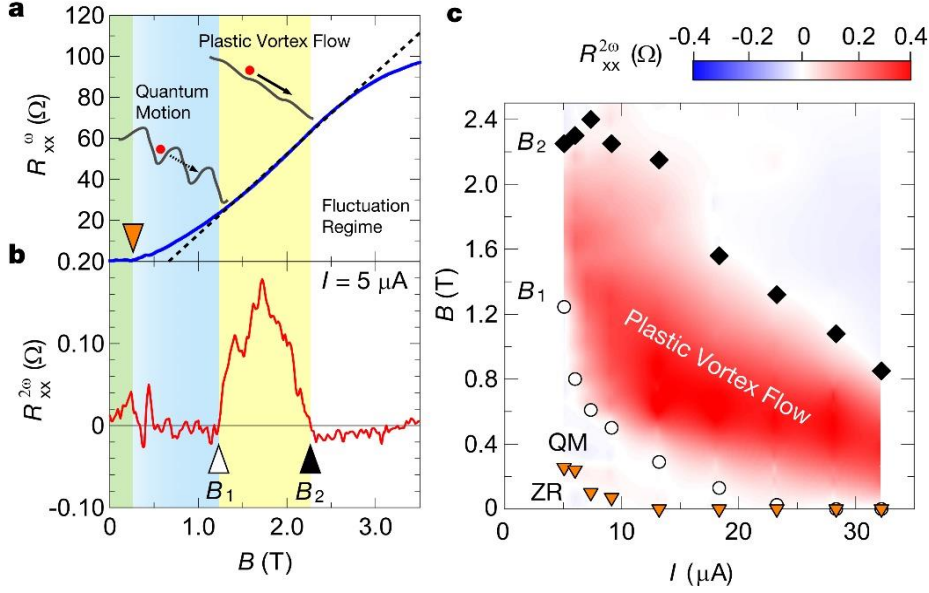

**Figure 4 | First and second harmonic magnetoresistance and dynamical vortex phase diagram.** (**a**) and (**b**), Magnetic field dependence of $R_{xx}^{\omega}$ (a) and $R_{xx}^{2\omega}$ (b) at $T$ = 2 K, $I$ = 5 µA. This particular sample shows the zero-$R_{xx}^{\omega}$. Zero-resistance (ZR) state up to the magnetic field indicated by the orange triangle. The black dashed line shows the *B*-linear dependence of $R_{xx}^{\omega}$ above 1.25 T. $R_{xx}^{2\omega}$ is significantly enhanced between $B_1$ and $B_2$ indicated by white and black triangles. The green, blue, yellow, and white backgrounds indicate the magnetic field regions, which are separated by the three triangles. Inset schematics display conceptual images of vortex motion in the corresponding magnetic field regions (see main text). (**c**), A dynamical vortex diagram in the *B-I* plane. While circles and black diamonds represent $B_1$ and $B_2$ extracted from Figs. 4b and 3e. Orange triangles show the magnetic field, where the ZR state disappears as extracted from the $R_{xx}^{\omega}$ - *B* curves for various values of *I*. Color mapping displays $R_{xx}^{2\omega}$ in the regime of 0 T < *B* < 2.5 T and 5 µA < *I* < 32 µA.

# Supplemental Material for
# Quantum and classical ratchet motions of vortices in a 2D trigonal superconductor

Yuki M. Itahashi, Yu Saito, Toshiya Ideue*, Tsutomu Nojima, Yoshihiro Iwasa*
*Correspondence to: ideue@ap.t.u-tokyo.ac.jp, iwasa@ap.t.u-tokyo.ac.jp

## I. AC measurement and definition of $\gamma$

Considering a system in the normal state with broken inversion symmetry and magnetic field $B$, an electrical resistance $R$ is generally written as

$$R = R_0(1+\gamma BI), \qquad (1)$$

where $R_0$ is a resistance under zero magnetic field and $\gamma$ the coefficient of nonreciprocal resistance in normal state[1–3]. It is noted that nonreciprocity emerges as a term in $R$ proportional to both $B$ and $I$.

However, when nonreciprocal transport stems from vortex flow rectified by asymmetric pinning potentials[4], $R$ is expressed by

$$R = R^{(1)}(1+\gamma'\hat{B}I), \qquad (2)$$

where $R^{(1)}$ is the resistance of first order, $\gamma'$ the coefficient of nonreciprocal resistance in this case and $\hat{B} = \dfrac{B}{|B|}$ the sign of magnetic field. It is noted that $R^{(1)}$ shows linear-like behavior but is even as a function of $B$. In this formula, the nonreciprocal term in $R$ is also proportional to $B$ as well as $I$. This definition is reasonable because nonreciprocity in vortex flow region is governed by the number of vortices, resulting in its linear dependence of $R^{(1)}$ on $B$. When the AC bias current (source-drain current) with a frequency of $\omega$ ($I = I_0 \cos\omega t$) is applied, the out-put voltage is expressed as follows:

$$V = R^{(1)}I_0 \cos\omega t + \frac{R^{(1)}\gamma'\hat{B}I_0^2}{2}(1+\cos 2\omega t). \qquad (3)$$

Both $\omega$ and $2\omega$ components appear in eq. (3). At that time, by extracting $\omega$ component ($R^\omega$), we obtain

$$R^\omega = R^{(1)}. \qquad (4)$$

On the other hand, by extracting $2\omega$ component ($R^{2\omega}$),

$$R^{2\omega} = \frac{1}{2}R^\omega \gamma' \hat{B} I_0 \qquad (5)$$

is obtained. Thus, to determine $\gamma'$ value, we defined $\gamma'$ as

$$\gamma' = \frac{2R^{2\omega}}{R^\omega \hat{B} I_0}. \qquad (6)$$

We further approximated $R^{2\omega}$ and $R^\omega$ in eq. (6) by the peak value ($R^{2\omega}_{\text{peak}}$) and the $R^\omega$ value at the peak position ($R^\omega_{\text{peak}}$), respectively. This definition is different from one used in previous study[5] (approximated as $\gamma = \dfrac{2R^{2\omega}_{\text{peak}}}{R^\omega_{\text{peak}} B_{\text{peak}} I_0}$, where $B_{\text{peak}}$ is the magnetic field at the peak position). In the previous study, we thought of an effect of paraconductivity on nonreciprocal transport and thus expanded the concept of nonreciprocal transport in normal state[1–3], where nonreciprocal resistance is proportional to both $B$ and $I$. However, in the present system, the contribution of not particles but vortices is dominant. Therefore, we use the definition of $\gamma'$ value shown above. We compare it with $\gamma$ used in previous study in Fig. S1. The sudden increase in $\gamma$ just below $T_c$ described by the paraconductivity effect doesn't exist in $\gamma'$ because of decrease in $B_{peak}$ as $T$ approaches to $T_c$. In the present case, i.e., low temperature region, where $R^\omega$ and $R^{2\omega}$ show linear dependence on $B$ because the characteristics of the system is governed by number of vortices, the definition of $\gamma'$ might be more plausible.

## II. Symmetrization and antisymmetrization

To analyze nonreciprocal transport, we symmetrized or antisymmetrized the raw data obtained in the present experiment. We measured the magnetoresistance for both positive and negative magnetic field ($R_{exp}(B)$) and extracted the odd or even component as a function of $B$ by procedures shown below.

To extract symmetrized and antisymmetrized ($R_{sym}(B)$ and $R_{asym}(B)$) component from the raw data, we calculated

$$R_{sym}(B) = \frac{R_{exp}(B) + R_{exp}(-B)}{2} \qquad (7)$$

and

$$R_{asym}(B) = \frac{R_{exp}(B) - R_{exp}(-B)}{2}, \qquad (8)$$

respectively. $R_{sym}(B)$ and $R_{asym}(B)$ are even and odd as a function of $B$, respectively. We adopt these symmetrization or antisymmetrization for $R^\omega$ and $R^{2\omega}$ in the main text. We provide typical raw data and antisymmetrized data in Fig. S2.

The linear background component observed in Fig. 2e in the main text can be attributed to the Hall effect originating from the nonlinear component of source-drain current ($I^{2\omega}$), namely a component which is proportional to the square of in-put AC current in the current flowing through the sample. This $I^{2\omega}$ possibly comes from geometrical asymmetry of source and drain electrodes in the sample. When out-of-plane magnetic field is applied, $I^{2\omega}$ generates the transverse second harmonic resistance in the similar manner as the normal Hall effect in the first harmonic component, which cannot be distinguished from the intrinsic nonreciprocal resistance in the normal state. However, the anomalous increase of 2ω component with the peak in the superconducting states cannot be explained by the geometrical asymmetry, and thus we can safely conclude that this behavior intrinsically arises from the crystal symmetry of $MoS_2$.

**Supplementary reference**

**Supplementary Figures**

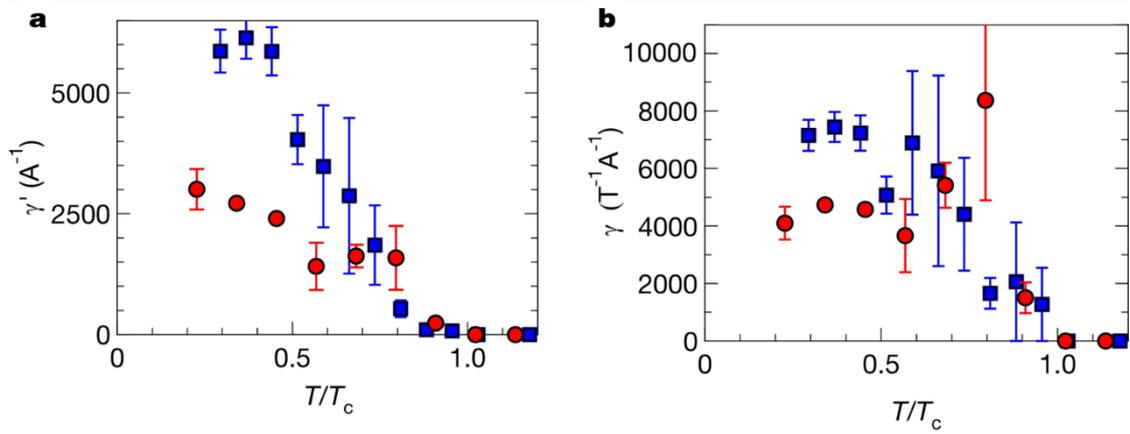

**Figure S1 | Comparison between $\gamma$ and $\gamma$'** (**a**) $\gamma$' as a function of $T/T_c$. $\gamma$' is defined in eq. (2). (**b**) $\gamma$ as a function of $T/T_c$. $\gamma$ is defined in eq. (1), which is used in previous studies. Red circles and blue squares represent $\gamma$ or $\gamma$' calculated from $R_{xx}^{2\omega}$ and $R_{xy}^{2\omega}$, respectively.

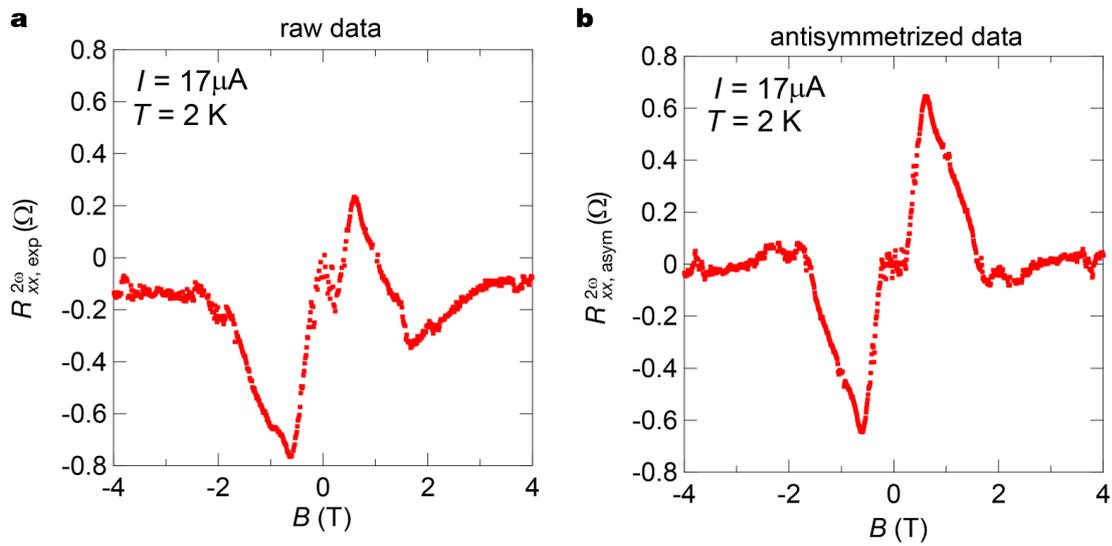

**Figure S2 | Raw and antisymmetrized second harmonic resistance** (**a**) and (**b**), (a) Raw and (b) antisymmetrized second harmonic resistance as a function of magnetic field at $I$ = 17 μA and $T$ = 2 K in sample 1.